\documentclass[a4paper]{article}

\usepackage{amsfonts}
\usepackage{amssymb}
\usepackage{bbold}
\usepackage{amsmath}
\usepackage{graphicx}
\usepackage{color}
\usepackage{fullpage}
\usepackage{pdfpages}

\usepackage[labelformat=empty]{caption}
\usepackage{epsfig}

\usepackage{tikz} 
\usepackage{eso-pic}

\newcommand{\gradientbox}[3]{%
  \begin{tikzpicture}
    \node[left color=#1,right color=#2] {#3};
  \end{tikzpicture}%
}

\newcommand{\Identity}{\mathbb{1}}

\begin{document}



# Topological quantum computing with a very noisy network and local error rates approaching one percent

Naomi H. Nickerson[1], Ying Li[2] & Simon C. Benjamin[2,3]

A scalable quantum computer could be built by networking together many simple processor cells, thus avoiding the need to create a single complex structure. The difficulty is that realistic quantum links are very error prone. A solution is for cells to repeatedly communicate with each other and so purify any imperfections; however prior studies suggest that the cells themselves must then have prohibitively low internal error rates. Here we describe a method by which even error-prone cells can perform purification: groups of cells generate shared resource states, which then enable stabilization of topologically encoded data. Given a realistically noisy network ($\geq 10\%$ error rate) we find that our protocol can succeed provided that intra-cell error rates for initialisation, state manipulation and measurement are below 0.82%. This level of fidelity is already achievable in several laboratory systems.

[1] Department of Physics, Imperial College London, Prince Consort Road, London SW7 2AZ, UK. [2] Centre for Quantum Technologies, National University of Singapore, 3 Science Drive 2, Singapore 117543. [3] Department of Materials, University of Oxford, Parks Road, Oxford OX1 3PH, UK. Correspondence and requests for materials should be addressed to S.C.B. (email: s.benjamin@qubit.org).

Topological codes are an elegant and practical method for representing information in a quantum computer. The units of information, or logical qubits, are encoded as collective states in an array of physical qubits[1]. By measuring stabilizers, that is, properties of nearby groups of physical qubits, we can detect errors as they arise. Moreover, with a suitable choice of stabilizer measurements we can even manipulate the encoded qubits to perform logical operations. The act of measuring stabilizers over the array thus constitutes a kind of 'pulse' for the computer—it is a fundamental repeating cycle and all higher functions are built upon it.

In the idealised case that the stabilizer measurements are performed perfectly, we can recover the logical qubit from a very corrupt topological memory (for example in the toric code ~19% of the physical qubits can be corrupt[2]). However, in a real device the stabilizer measurements will sometimes introduce errors, due to failures while initialising ancillas, performing gate operations or measurements. Will the act of measuring stabilizers still 'do more good than harm' and protect the encoded information? This depends on the frequency of such errors; estimates of the threshold error rate are 0.75–1.4% depending on the model[3–5].

These numbers are applicable to a 'monolithic' architecture, but for many systems a network architecture may be more scalable (Fig. 1). Links will be error prone, but cells can interact repeatedly and purify the results to remove errors[6,7]. Thus one can realise a 'noisy network' (NN) paradigm, also called distributed[8,9] or modular[10]. However prior analyses of this approach have indicated a serious drawback. The need to perform purification means a cell's own internal error rate must be very low; of the 0.1% order[6–10]. Given that a real machine should operate well below threshold, such a demand may be prohibitive. It is timely to seek a less demanding approach, given that entanglement over a remote link has now been demonstrated with NV centres[11], adding diversity to the previous atomic experiments[12].

Here we describe a new NN protocol which achieves a threshold for the intra-cell operations which is far higher, and comparable to current estimates for tolerable noise in monolithic architectures. Our cells collectively represent logical qubits according to the 2D toric code (one of the simplest and most robust topological surface codes). This code is stabilized by repeatedly measuring the parity of groups of four qubits in either the Z or the X basis. In our approach each elementary 'data qubit' of the code resides in a separate cell, alongside a few dedicated ancilla qubits. We implement stabilizer measurements by directly generating a GHZ state shared across the ancillas in four cells, before consuming this resource in a single step to measure the stabilizer on the four data qubits. This procedure is in the spirit of the bandaid protocol[13] and contrasts with the standard approach of performing a sequence of two-qubit gates between the four data qubits and a fifth auxiliary qubit (which, in the network picture, would require its own cell). By directly generating the GHZ state over the network we remove the need for the auxiliary unit and more importantly we can considerably reduce the accumulation of errors.

## Results

**Protocol.** In this paper we consider three variants of our noisy network protocol. In all cases, one qubit in each cell is designated the data qubit and the others are ancillas that will be initialised, processed and measured during each stabilizer evaluation process. Figure 2 defines two of the variants: if we neglect the elements enclosed in dashed lines then we have the compact EXPEDIENT protocol, meanwhile applying all steps yields the STRINGENT alternative. The former has the advantage that it requires less time to perform. Presently we also introduce a further protocol called STRINGENT+ which requires an additional qubit in each cell. Establishing the performance of these three variants is the principle aim of this paper. The results are detailed presently; in summary, setting the network error rate to 10% one finds threshold local error rates of about 0.6, 0.775 and 0.825% for EXPEDIENT, STRINGENT and STRINGENT+ respectively. Optionally we can trade local error rate for still greater network noise tolerance, see Supplementary Note 1. These thresholds are considerably higher than those in prior work; for example in a recent paper[8] we applied an optimised purification protocol[14] to the case of a network with three qubits per node, finding a threshold local error rate in the region of 0.1%. The improved rates seen in the present paper result from generating and purifying a GHZ resource state over the network, followed by a one-step stabilizer measurement. It seems that the additional storage in this new method requires at least four qubits per cell.

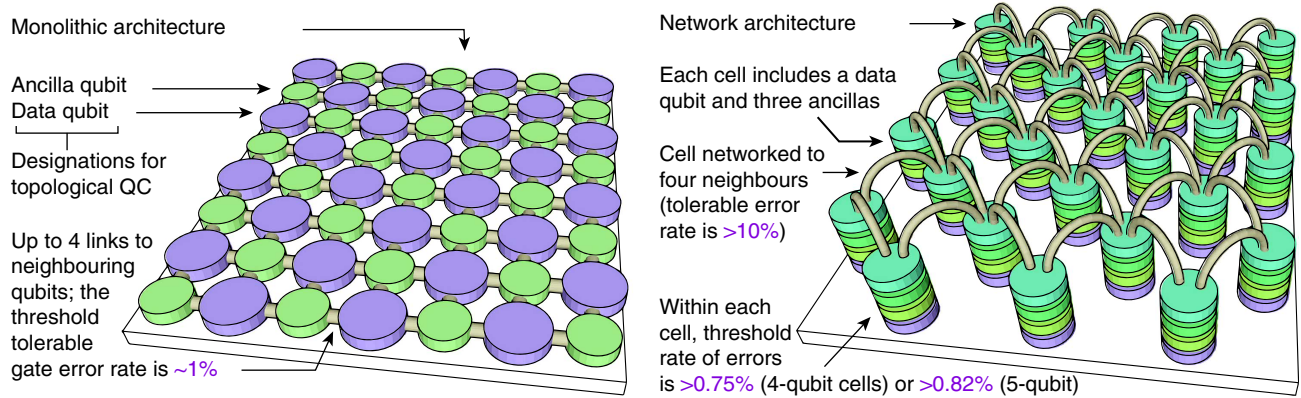

**Figure 1 | Quantum architectures.** Left: a monolithic grid of qubits with neighbours directly connected to enable high fidelity two-qubit operations. (layout from ref. 5.) Such a structure is a plausible goal for some systems, for example, specific superconducting devices[23]. Right: For other nascent quantum technologies the network paradigm is appropriate. A single nitrogen-vacancy (NV) centre in diamond, with its electron spin and associated nuclear spin(s)[24], would constitute a cell. A small ion trap holding a modest number of ions[10,25] is another example. Noisy network links with error rates ≥ 10% are acceptable. For photonic links this goal is realistic given imperfections like photon loss and instabilities in path lengths or interaction strengths. Similarly, with solid state 'wires' formed by spin chains[26], noisy entanglement distribution of this kind is a reasonable goal[27].

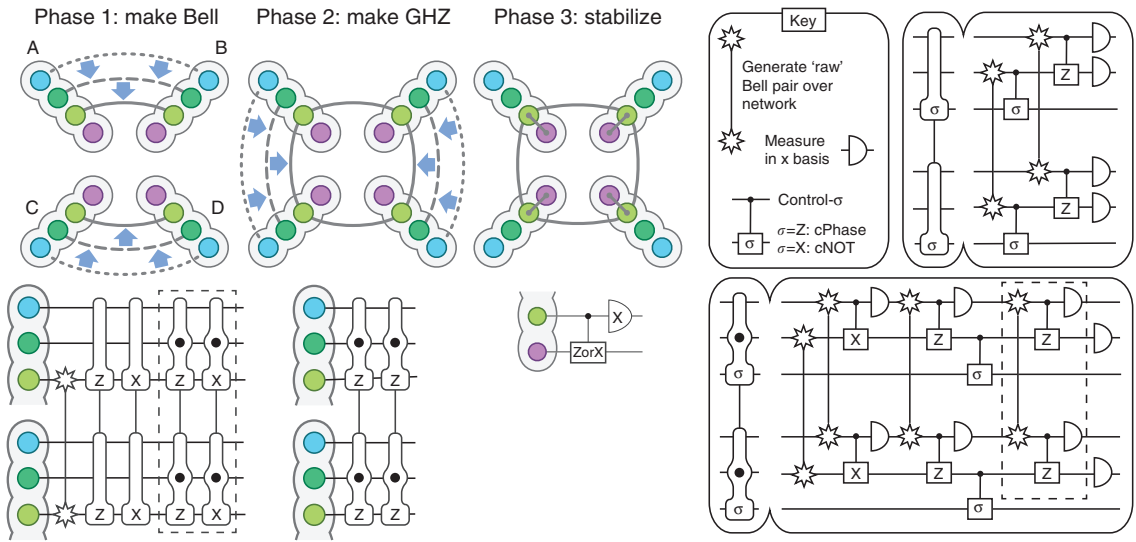

**Figure 2 | Protocols for stabilizer measurement.** The three phases for generating and consuming a 4-cell GHZ state among the ancilla qubits (blue-to-green) in order to perform a stabilizer measurement on the data qubits (purple). Phase (1): use purification to create a high quality Bell pair shared between cell A and cell B, while in parallel doing the same thing with cells C and D. (2) Fusion operations, A to C and B to D, create a high fidelity GHZ state. (3) Finally use the GHZ state to perform a one-step stabilizer operation. The parity of the four measured classical bits is also the parity of the stabilizer operation we have performed on the data qubits. Two dashed regions indicate operations that are part of the STRINGENT protocol; omitting them yields the EXPEDIENT alternative.

We assume the network channel between two cells can create shared, noisy Bell pairs in the Werner form

$$\rho_{\text{raw}} = \left(1 - \frac{4p_n}{3}\right) |\Phi^+\rangle\langle\Phi^+| + \frac{p_n}{3}\mathbb{1} \qquad (1)$$

where $|\Phi^+\rangle = (|00\rangle + |11\rangle)/\sqrt{2}$. Throughout this paper unless otherwise stated we take the network error probability to be $p_n = 0.1$. This 'raw' Bell generation is the sole operation that occurs over the network. We additionally require only three intra-cell operations: controlled-$Z$ (that is, the two-qubit phase gate), controlled-$X$ (that is, the control-NOT gate) and single qubit measurement in the $X$-basis. The two qubit gates are modelled in the standard way: an ideal operation followed by the trace-preserving noise process

$$N(\rho) = (1 - p_g)\rho + \frac{p_g}{15}\sum_{A,B}(A \otimes B)\rho(A \otimes B)^\dagger \qquad (2)$$

where operator $A \in \{\mathbb{1}, \sigma_x, \sigma_y, \sigma_z\}$ acts on the first qubit, and similarly $B$ acts on the second qubit, but we exclude the case $\mathbb{1} \otimes \mathbb{1}$ from the sum. Meanwhile noisy measurement is modelled by perfect measurement preceded by inversion of the state with the probability $p_m$. Phases 1 and 2 in Fig. 2 are postselective: whenever we measure a qubit, one outcome indicates 'continue' and the other indicates that we must recreate the corresponding resource (the desired outcome is that which we would obtain, were the noise parameters set to zero). We note that in the circuits developed here we adopt and extend the 'double selection' concept introduced by Fujii and Yamamoto in ref. 15. Resource overhead can be quite modest: EXPEDIENT minimises the number of raw pairs required and thus the overall time requirement. On average it requires fewer than twice the number of pairs consumed in the ideal case that all measurements 'succeed', see Supplementary Note 2.

**Performance evaluation.** Having specified the stabilizer measurement protocols, we must determine their real effect given the various error rates $p_n$, $p_g$ and $p_m$. It is convenient derive a single

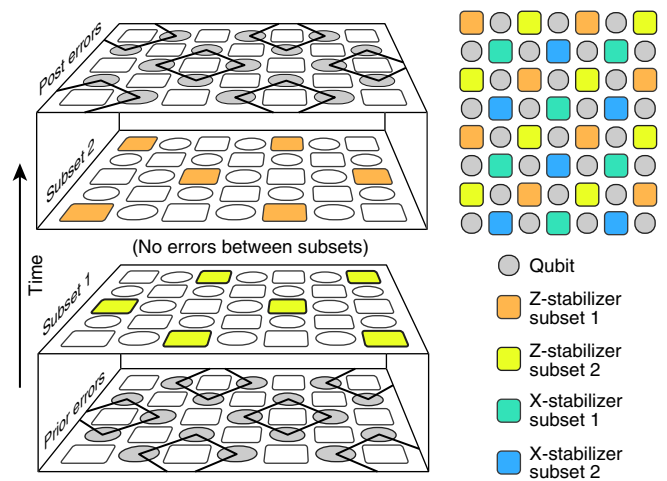

**Figure 3 | Scheduling stabilizer operations.** The right side graphic shows the standard arrangement of one complete stabilizer cycle, involving $Z$ and $X$ projectors (square symbols indicate that the four surrounding data qubits are to be stabilized). Because a given cell can only be involved in generating one GHZ resource at a given time, each of these two stabilizer types must be broken into two subsets; see main figure. Fortunately in our stabilizer superoperator we can commute projectors and errors so as to expel errors from the intervening time between subsets, so allowing them to merge.

superoperator describing the action of the entire stabilizer protocol on the four data qubits. This is described in the Methods section below. Given this operator, together with a suitable scheduling scheme as shown in Fig. 3, we proceed to determine thresholds by intensive numerical modelling. Our model introduces errors randomly but with precisely the correct correlation rates. We record the (noisy) stabilizer outcomes and subsequently employ Edmonds' minimum weight perfect matching algorithm[16], implemented as described in ref. 17, to pair and resolve

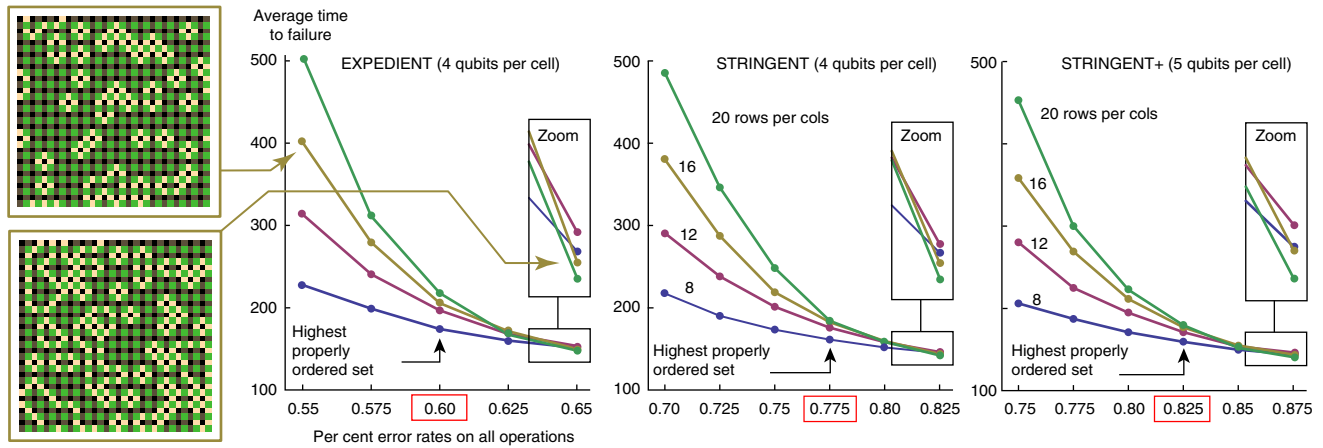

**Figure 4 | Performance of the EXPEDIENT and STRINGENT protocols.** We employ a toric code with $n$ rows × $n$ columns of data qubits ($2n^2$ in total). A given numerical experiment is a simulation of 100 complete stabilizer measurement cycles on an initially perfect array, after which we attempt to decode a $Z$ measurement of the stored qubit. The result is either a success or a failure; for each data point we perform at least 10,000 experiments to determine the fail probability, and reciprocate this to infer an expected time to failure. Network error rates are 10% in all cases; we set intra-cell gate and measurement error rates equal, $p_m = p_g$, and plot this on the horizontal axes. For low error rates the system's performance improves with increasing array size. As the error rates pass the threshold this property fails. Insets: typical final states of the toroidal array after error correction. Yellow squares are flipped qubits, green squares indicate the pattern of Z-stabilizers. Closed loops are successful error corrections; while both arrays are therefore successfully corrected, it is visually apparent that the above-threshold case is liable to long paths.

stabilizer flips in the standard way (see for example, ref. 5). This technique allows us to establish the threshold for successful protection of quantum information by simulating networks of various sizes. If an increase of the network size allows us to protect a unit of quantum information more successfully, then we are below threshold. Conversely if increasing the network size makes things worse, then in effect the stabilizers are introducing more noise than they remove and we are above threshold. The results of a large number of such simulations are shown in Fig. 4 (and equivalent data for the monolithic case are shown in Supplementary Fig. S1). In these Figures we sweep the local error rate to determine the threshold; equivalently one can sweep the network error rate as shown in Supplementary Fig. S2. We see thresholds for EXPEDIENT and for STRINGENT of over 0.6 and 0.77% respectively. The third graph shows the performance of a five-qubit-per-cell variation of STRINGENT which we describe in more detail in the Supplementary Note 1. It achieves a threshold in excess of 0.82% by employing an additional filter such that most stabilizer measurements are improved while a known minority become more noisy, and this classical information is fed into an enhanced Edmonds' algorithm.

**External decoherence.** We have yet to consider general decoherence caused by the external environment during a stabilizer protocol. Fortunately many of the systems most relevant to the NN approach have excellent low-noise 'memory qubits' available. For example in NV centres at room temperature, nuclear spins can retain coherence for the order of a second; and for impurities in silicon the record is several minutes[18,19]. We would naturally use such spins for our data qubits and for the innermost ancilla qubits, that is, those bearing the GHZ-state in Fig. 2. In Supplementary Note 3 we summarise the relevant control and measurement timescales from recent NV centre and atomic experiments, and thus estimate the probability of an environmentally induced error during a full stabilizer protocol. We find that such error probabilities are small compared to the error rates $p_g$ and $p_m$ that we have already considered.

**Computation versus qubit storage.** The results described here are thresholds for successful protection of encoded logical qubits. We do not explicitly simulate the act of performing a computation involving two or more logical qubits. However it is generally accepted that the threshold for full computation will be very similar to that for successful information storage, because the stabilizer measurements required for simple preservation of the encoded state are very similar to those required for computation[3,20]. In both cases, the standard four-qubit stabilizer is overwhelmingly the most common measurement. For computation it is necessary to evaluate additional forms of stabilizer, for example a stabilizer measurement corresponding to the ($X$ or $Z$-basis) parity of three, rather than four, qubits. This applies both in the approach of braiding defects[3] and the alternative idea of performing 'lattice surgery'[20]. However these alternative stabilizers are a minority, being required at boundaries while the four-qubit standard stabilizer still forms the bulk of operations. Moreover both for monolithic architectures and the noisy network paradigm, the three-qubit stabilizer will in fact have a somewhat lower error rate since it requires a less complex protocol.

**Discussion**

We have described an approach to 'noisy network' quantum computing, where many simple modules or cells are connected with error prone links. We show that relatively high rates of error within each cell can be tolerated, while the links between the cells can be very error prone (10% or more). This work therefore largely closes the gap between error tolerance in networks versus monolithic architectures. We hope this result will encourage the several emerging technologies for which networks are the natural route to scalability; these include ion traps and NV centres (linked either optically or via spin chains).

Throughout this study we have assumed that all forms of error are equally likely; in reality, in a given physical system some errors may be more prevalent. For example phase errors on the network channel might be more common than flip errors, and similarly the local gates within cells may suffer specific kinds

of noise. Any such bias in error occurrences is 'good news' in that it can potentially be exploited by adapting our protocols, and in this way the error thresholds might be further increased.

## Methods

**Derivation of superoperator.** In the following we write $M$ to stand for the 'odd' or 'even' reported outcome of the stabilizer protocol (that is, the parity of the four measured qubits in Phase 3 of Fig. 2). If density matrix $\rho$ represents the state of all data qubits prior to the evaluation of our stabilizer, then measurement outcome $M$ and the corresponding state $S^M(\rho) = P^M(\rho)/\text{Tr}[P^M(\rho)]$ will occur with probability $\text{Tr}[P^M(\rho)]$, for some projective operator $P^M(\cdot)$. Now stabilizing the toric code involves two types of operation, measuring the $X$ and the $Z$—stabilizers. Suppose we are performing a $Z$ stabilizer. If our protocol could act perfectly, then for example the even projector would be $P^{even}_{ideal}(\rho) = \sum |i\rangle\langle i| \rho |j\rangle\langle j|$, where the sum is over all states $|i\rangle$, $|j\rangle$ with definite even parity in the $Z$-basis: $|0000\rangle$, $|0011\rangle$, etc. Analogously the ideal odd $Z$-stabilizer sums over states of definite odd parity, and meanwhile for $X$ stabilizers the ideal projectors refer to states of definite parity in the $X$-basis. In reality our imperfect operations result in projectors of the form

$$P^M_{real}(\rho) = \sum_e a^M_e E_e P^M_{ideal}(\rho) E_e^\dagger + b^M_e E_e P^{\bar{M}}_{ideal}(\rho) E_e^\dagger$$

with $E_e = (ABCD)_e$ and $\{A, B, C, D\} \in \{\mathbb{1}, \sigma_x, \sigma_y, \sigma_z\}$.

Here the four operators making up $E$ are understood to act on data qubits 1 to 4 respectively, and index $e$ runs over all their combinations. The symbol $\bar{M}$ represents the compliment of $M$, that is, 'odd' for $M$ = 'even' and vice versa. We see that this real projector is made up of a mix of the correct and incorrect projectors together with possible Pauli errors; the various weights $a$ and $b$ capture their relative significance.

Given the underlying error rates $p_n$, $p_g$ and $p_m$ we can employ the Choi-Jamiolkowski isomorphism to find the corresponding weights $\{a, b\}$ in our superoperator. Incorporated in this process is a 'twirling' operation; this is explained in the Supplementary Methods along with examples of the resulting weights. The largest contributor to $P^M_{real}(\rho)$ is found to be the reported parity projection, that is, for $M$ = 'even', the largest of all weights $\{a, b\}$ is that associated with $E = (\mathbb{1}\mathbb{1}\mathbb{1}\mathbb{1})$ and $P^{even}_{ideal}(\rho)$. The next largest term will be the pure 'wrong' projection, i.e. the combination of $E = (\mathbb{1}\mathbb{1}\mathbb{1}\mathbb{1})$ and $P^{odd}_{ideal}(\rho)$. This form of error is relatively easy for the toric code to handle, and we have deliberately favoured it over other error types by minimising the $\sigma_x$ and $\sigma_y$ errors, rather than $\sigma_z$, in our stabilizer measurement protocols. The remaining terms correspond to Pauli errors in combination with either the correct or the incorrect projectors; for example $E = (\sigma_x\sigma_z\mathbb{1}\mathbb{1})$ is an erroneous flip on data qubit 1 simultaneous with a phase error on qubit 2.

**Decoding algorithm.** Edmonds' algorithm was selected, having been well studied in the context of noisy stabilizer measurements; in Supplementary Fig. S2 we apply our same model to monolithic architecture, obtaining a threshold in the region of 0.9%, generally consistent with prior studies[5]. For the idealised case of perfect (noiseless) stabilizer measurements other algorithms have been developed which may eventually offer advantages for noisy stabilizers as well[2,21,22].

## Acknowledgements

## Author contributions

N.H.N., Y.L. and S.C.B. designed the protocols. N.H.N. and S.C.B. performed circuit simulations. Y.L. and S.C.B. calculated thresholds. S.C.B. wrote the paper.

## Additional information

**Supplementary Information** accompanies this paper at http://www.nature.com/naturecommunications

**Competing financial interests:** The authors declare no competing financial interests.

**Reprints and permission** information is available online at http://npg.nature.com/reprintsandpermissions/

**How to cite this article:** Nickerson, N.H. *et al.* Topological quantum computing with a very noisy network and local error rates approaching one percent. *Nat. Commun.* 4:1756 doi: 10.1038/ncomms2773 (2013).

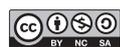



\AddToShipoutPicture{%
  \AtPageLowerLeft{%
    \rotatebox{90}{
      \gradientbox{blue!20}{white}{%
        \begin{minipage}{\paperheight}%
          \hspace*{ \stretch{1} }{\Large Supplementary Information} \hspace*{ \stretch{1} }
        \end{minipage}%
      }
    }%
  }%
}

\bigskip
\bigskip


\noindent {\noindent {\bf \Large  Supplementary Figure S1: Comparison to the monolithic case}

\begin{figure}[h!]
\centering
\includegraphics[width=0.9\columnwidth]{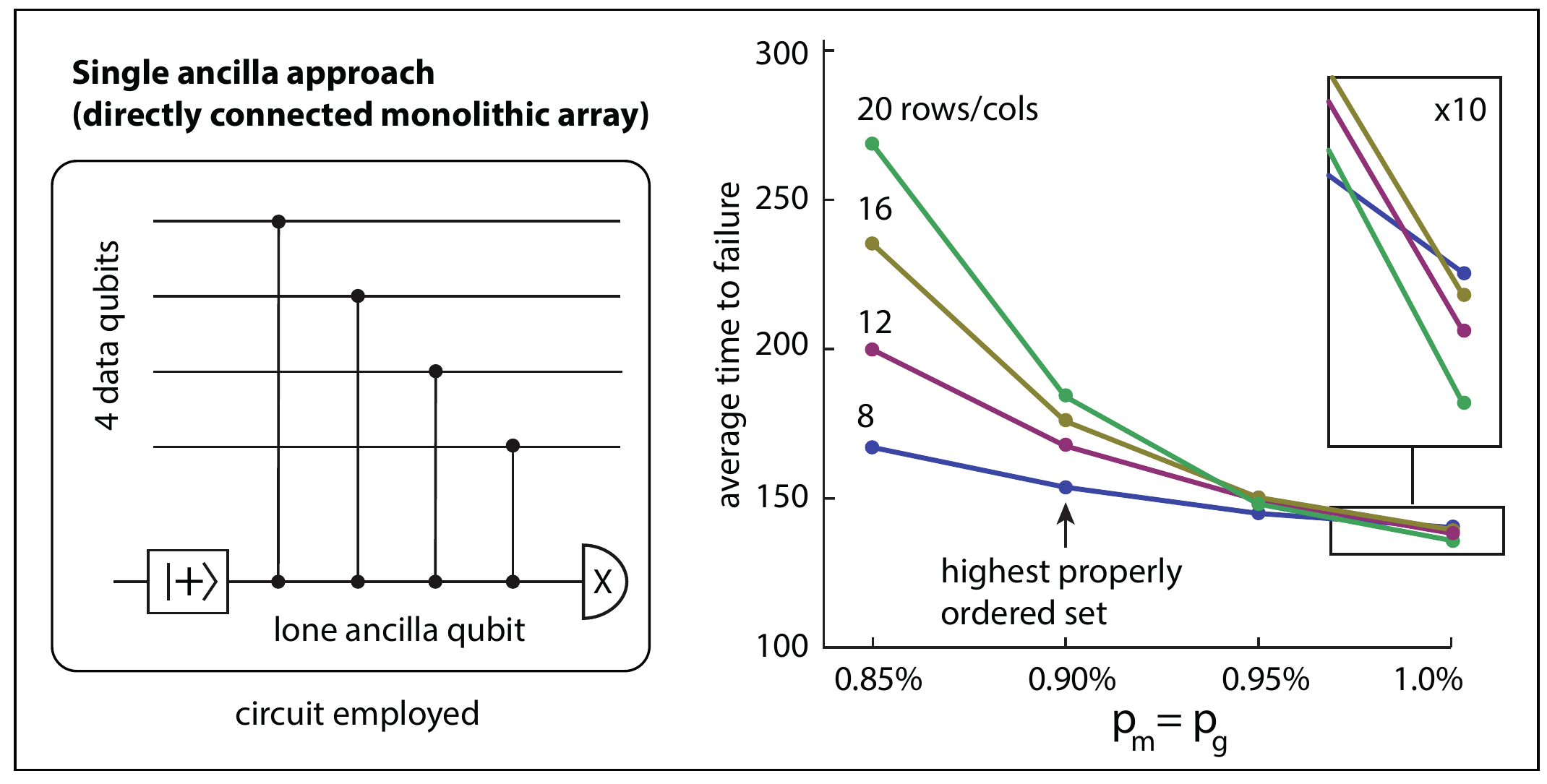}
\caption{ Performance of the monolithic architecture. This Figure is to be compared with the graphs in Fig. 4 of the main paper. In order to make a clear comparison with the threshold that can be achieved with the monolithic architecture, we applied our same superoperator description and numerical simulation to this case. (Of course there is no meaning for $p_n$, the network error rate, in a monolithic architecture). The figure shows the circuit we used for the case of a $Z$ stabilizer. Note that this approach requires initialisation of the single shared auxiliary qubit; we used the value of $p_m$ as the fidelity of this initialisation.
One sees that the threshold is between $0.9\%$ and $0.95\%$ and is therefore appreciably higher than that obtained by {\small STRINGENT} whilst in the same ``ball park''. As noted below, the addition of another ancilla per cell can further close this gap.
}
\label{data}
\end{figure}


\newpage

\noindent {\noindent {\bf \Large   Supplementary Figure S2: Sweeping the network error rate\\}

\begin{figure}[h]
\centering
\includegraphics[width=0.55\columnwidth]{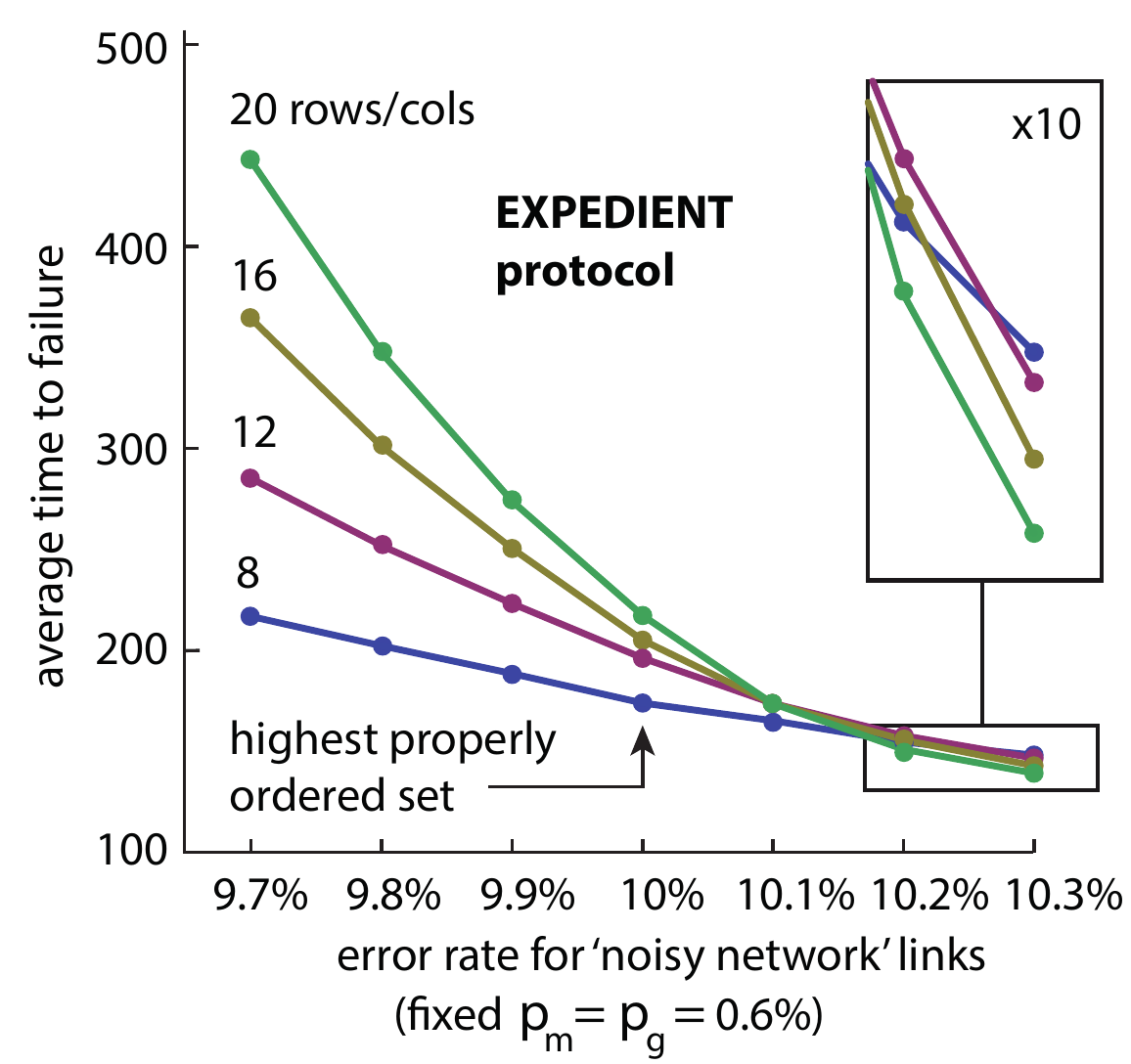}
\caption{Varying network error rate with fixed local error rates. In the main paper we set the network error rate to $10\%$ and consider a range of local (intra-cell) error rates in order to determine the threshold. We find thresholds in the range $0.6\%$ to $0.82\%$ depending on the details of the protocol. We have also performed a complimentary series of simulations where we fix the local error rate and sweep the network error rate. Supplementary  Figure 2 shows the corresponding data for the case of the {\small EXPEDIENT} protocol. Fixing the local error rates $p_m=p_g=0.6$, our sweep of the network error rate reveals a threshold in the $10.0\%$ to $10.1\%$ range, consistent with the calculations in the main paper. 
}
\label{data}
\end{figure}

\bigskip

\noindent {\noindent {\bf \Large   Supplementary Tables\\}

\begin{table}[!h]
\centering
\begin{tabular}{|c|c|c|c|c|}
\hline 
Level, $L$ &  & Time steps, $t_{L}$ & $p_{success,L}$ & FRL\tabularnewline
\hline 
\hline 
1 & Round one Bell pair production & 7 & 0.7346 & 1\tabularnewline
\hline 
2 & Round two Bell pair production & 6 & 0.7506 & 1\tabularnewline
\hline 
3 & Round one single selection & 4 & 0.8619 & 3\tabularnewline
\hline 
4 & Round two single selection & 3 & 0.8550 & 3\tabularnewline
\hline 
5 & Make GHZ & 2 & 0.8651 & 1\tabularnewline
\hline 
6 & Round one single selection & 4 & 0.8619 & 6\tabularnewline
\hline 
7 & Round two single selection & 3 & 0.8550 & 7\tabularnewline
\hline 
8 & Check GHZ & 2 & 0.8654 & 1\tabularnewline
\hline 
9 & Measure stabilizer & 2 & - & -\tabularnewline
\hline 
\end{tabular}
\caption{\label{tab:table} {\bf Supplementary Table S1:} Listing of probabilities used in modelling the number of steps required for {\small EXPEDIENT}. FRL stands for Failure Reset Level. Please see Supplementary Note 2 for further details.}
\end{table}

\bigskip

\begin{table}[!h]
\centering
\begin{tabular}{|c|c|c|c|c|}
\hline 
Level, $L$ &  & Time steps, $t_{L}$ & $p_{success,L}$ & FRL\tabularnewline
\hline 
\hline 
1 & Bell pair production & 7 & 0.7277 & 1\tabularnewline
\hline 
2 & Bell pair check 1 & 6 & 0.7429 & 1\tabularnewline
\hline 
3 & Round one single selection & 4 & 0.8586 & 3\tabularnewline
\hline 
4 & Round two single selection & 3 & 0.8509 & 3\tabularnewline
\hline 
5 & Bell pair check 2 & 5 & 0.8019 & 1\tabularnewline
\hline 
6 & Round one single selection & 4 & 0.8586 & 6\tabularnewline
\hline 
7 & Round two single selection & 3 & 0.8509 & 6\tabularnewline
\hline 
8 & Bell pair check 3 & 5 & 0.8043 & 1\tabularnewline
\hline 
9 & Round one single selection & 4 & 0.8586 & 9\tabularnewline
\hline 
10 & Round two single selection & 3 & 0.8509 & 9\tabularnewline
\hline 
11 & Make GHZ & 5 & 0.6588 & 1\tabularnewline
\hline 
12 & Round one single selection & 4 & 0.8586 & 12\tabularnewline
\hline 
13 & Round two single selection & 3 & 0.8509 & 12\tabularnewline
\hline 
14 & Check GHZ & 5 & 0.6454 & 1\tabularnewline
\hline 
15 & Measure stabilizer & 2 & - & -\tabularnewline
\hline 
\end{tabular}
\caption{\label{tab:stringentTable} {\bf Supplementary Table S2:} As Supplementary Table S1, but now for {\small STRINGENT} protocol with $p_{g}=p_{m}=0.75\%$,
$p_{n}=10\%$. Please see Supplementary Note 2 for further details. \\ \\ \\ }
\end{table}

\bigskip

\begin{table}[!h]
\centering
\begin{tabular}{|c|c|c|c|c|}
\hline 
Label for $Z$ proj& Label for $X$ proj&  Case 1 &  Case 2 & Case 3\tabularnewline
\hline 
\hline 
$A_\Identity$ &	$A_\Identity$&	0.9117 &		0.928& 		0.951\tabularnewline
\hline 
$A_Z$ & 		$A_X$ & 		0.00681&		0.00675&		0.00470\tabularnewline
\hline 
$A_X$ & 		$A_Z$ & 		0.00314&		0.00391&		0.00352\tabularnewline
\hline 
$A_Y$ &		$A_Y$ & 		0.00314&		0.00391&		0.00352\tabularnewline
\hline 
$A_{ZZ}$ &	$A_{XX}$ & 	0.000182&	0.000187&	0.00119\tabularnewline
\hline 
$A_{XX}$& 	$A_{ZZ}$ & 	0.00000758&	0.00001182&	0.0000128\tabularnewline
\hline 
$A_{YY}$ & 	$A_{YY}$ & 	0.00000758&	0.00001182&	0.0000128\tabularnewline
\hline 
$A_{XZ}$ & 	$A_{XZ}$ & 	0.0000336 &	0.0000414&	0.00121\tabularnewline
\hline 
$A_{YZ}$ & 	$A_{XY}$ & 	0.0000336&	0.0000414&	0.00121\tabularnewline
\hline 
$A_{XY}$ & 	$A_{YZ}$ & 	0.0000152&	0.0000236&	0.0000256\tabularnewline
\hline 
$B_\Identity$ &	$A_\Identity$&	0.0617&		0.0424& 		0.0178\tabularnewline
\hline 
$B_Z$ & 		$B_X$ & 		0.00674&		0.00665&		0.00470\tabularnewline
\hline 
$B_X$ & 		$B_Z$ & 		0.00314&		0.00391&		0.00352\tabularnewline
\hline 
$B_Y$ &		$B_Y$ & 		0.00314&		0.00391&		0.00352\tabularnewline
\hline 
$B_{ZZ}$ &	$B_{XX}$ & 	0.000127&	0.0000849&	0.00119\tabularnewline
\hline 
$B_{XX}$& 	$B_{ZZ}$ & 	0.00000758&	0.00001182&	0.0000128\tabularnewline
\hline 
$B_{YY}$ & 	$B_{YY}$ & 	0.00000758&	0.00001182&	0.0000128\tabularnewline
\hline 
$B_{XZ}$ & 	$B_{XZ}$ & 	0.0000336 &	0.0000414&	0.00121\tabularnewline
\hline 
$B_{YZ}$ & 	$B_{XY}$ & 	0.0000336&	0.0000414&	0.00121\tabularnewline
\hline 
$B_{XY}$ & 	$B_{YZ}$ & 	0.0000152&	0.0000236&	0.0000256\tabularnewline
\hline 
\end{tabular}
\caption{\label{tab:tableSO} {\bf Supplementary Table S3:} Listing of weighting coefficients for three examples superoperators, as explained in the Supplementary Methods.  \\}
\end{table}

\newpage

\noindent {\noindent {\bf \Large   Supplementary Note 1: Five qubits per cell\\}

The protocols {\small EXPEDIENT} and {\small STRINGENT} use a total of four qubits per cell of the network (one data qubit, three ancillas). If one adds further ancilla(s) then the performance improves. While it is beyond the scope of the present paper to seek optimal protocols that exploit five (or more) qubits, we can easily modify our approaches to make some use of the additional resource to tolerate more severe network error rates. For example, wherever our  original 4-qubit {\small STRINGENT} protocol calls for a ``raw'' Bell pair to be created on a given ancilla pair we can instead insert a small circuit that creates raw pairs using the additional ancillas and purifies them. Similarly one can replace ``single selection'' purification (two tiers of ancilla yield an improved pair on the higher tier) with ``double selection'' (three tiers of ancillas yield a significantly improved pair on the highest tier). Adopting these rather naive modifications we immediately find that the network noise is tolerable at the $p_n=0.2$ level (rather than $p_n=0.1$) for the same threshold of $0.77\%$ local errors. 

Alternatively we can hold the network errors constant and improve our tolerance of intra-cell errors. Data for this case are shown in the third panel of Fig.~4 in the main paper, labelled ``{\small STRINGENT+}''. We have introduced one further enhancement: We replace the usual Phase 3 of the  {\small STRINGENT} protocol i.e. the steps where the GHZ resource would simply be coupled to the data qubits and then measured out. Instead we filter the GHZ {\em after} coupling it to the data qubits, and if this filter fails we {\em abort} the protocol by measuring the GHZ in the $Z$-basis (this therefore requires the addition of $Z$-basis measurement to our set of allowed primitive operations). If we have aborted, we then perform a whole new round of stabilization, this time without the filter so that the protocol will certainly complete. This procedure leaves us knowing the stabilizer outcome {\em and} some additional classical information about exactly what steps occurred. Specifically there are $3$ distinct cases: (a) The filter was successfully passed; this is the best case and results in lower error rates on the data qubits -- it is about $92\%$ of cases in for the parameter range we consider. (b) The filter failed, but on measuring the GHZ in the $Z$-basis it was found to be in a correct state (e.g. $0000$ or $1111$). This occurs about $4\%$ of the time. It is another ``good'' case in that there is a low chance of errors having reached the data qubits, so that the second round can perform normally. (c) The filter failed, and on measuring on the GHZ in the $Z$-basis it was found to be in an incorrect state (e.g. $0001$). This occurs about $4\%$ of the time and it is the ``bad'' case; in this event there is very likely to be an error on the data qubits. Now if we make no use of the classical information and merely ``forget it'' then the net effect of this protocol is to make things worse versus a simple one-round use of {\small STRINGENT} -- this is not surprising since the overall risk of an error is not reduced (it is increased slightly due to the extra steps). However if we modify our Edmonds matching algorithm to use the classical information, specifically to favour paths that are consistent with errors occurring where ``bad'' stabilizers took place, then the threshold improves (this is the case shown in the third panel of Fig.~4). In effect we trade a small amount of increased error risk for a significant amount of classical knowledge: the $4\%$ of ``bad'' stabilizers account for about half of all errors entering the system.

\bigskip

\noindent {\noindent {\bf \Large   Supplementary Note 2: Time costs and memory errors\\}

Since our stabilizer measurement protocols are considerably longer than the equivalent procedures for a monolithic architecture, and indeed our approach is post-selective and therefore of uncertain duration, it is important to assess the potential impact of memory errors. In the main paper we assert that memory errors should have negligible impact, if our cells can employ qubits at least as good as those demonstrated in the { Science} papers of  Maurer {\em et al} and Steger {\em et al} (Refs.~[23] and [24] of the main paper). To substantiate this we need to estimate the duration of the protocols. In the following we assign one ``time step'' to any elementary operation in our protocol, whether a gate, a raw Bell creation or a measurement. We assume that the operations within a cell must be strictly sequential. Because we are evaluating an entire `sheet' of stabilizers in parallel across a large array of many cells (see Fig.~3 of the main paper), we need to be concerned not merely with the average time that a stabilizer protocol might take, but rather with the the time required for a given target proportion of all stabilizers to succeed. 

In Supplementary Table S1 and Supplementary Table S2 we give data for both {\small EXPEDIENT} and {\small STRINGENT}. In the following analysis we will focus on the former since it is the protocol designed for use when memory errors are an issue. 
The process of building a GHZ state can be separated into distinct
sections, each of which is terminated by a measurement, which, if
it results in the `wrong' outcome, will reset the process to an earlier
stage. Each of these measurement outcomes has an associated probability
of success, $p_{L}$, where the index $L$ denotes the level. For
the EXPEDIENT protocol operating at error rates of $p_{g}=p_{m}=0.6\%$
and $p_{n}=10\%$, these levels and their probabilities are given
in Supplementary Table S1. It should be noted that levels 1 and 2
utilise two cells (and 3,4 and 6,7) are run twice in parallel accross
four cells, and consequently the longer of the two times will determine
when the process may proceed to the next level. 

The GHZ production process is probabilistic, with a minimum duration
of $\sum_{L}t_{L}=33$, which occurs with a probability $\prod_{L}p_{L}=0.2242$.
Using the parameters in Supplementary Table~S1 the general process was simulated; 100,000
samples were generated and used to estimate the parameters of the
resulting distribution. This found the expected completion time for a given stabilizer measurement to be
to be 68.2 time steps, while 50\% of operations were completed after
57 time steps, 95\% after 138 time steps, 99\% after 195 time steps
and 99.9\% after 278 time steps.

{\color{black} Our topological code simulations indicate that if we wait for $99\%$ of stabilizers to evaluate, and abandon the remaining $1\%$, then there is negligible impact on the threshold. (A stabilizer measurement that fails to report is less damaging than a stabilizer measurement that performs a `wrong' projection; moreover such stabilizers will not introduce errors since the ancilla GHZ state is never created and coupled to the data qubits.) Therefore we simply take the expected time for $99\%$ of stabilizer protocols to complete as the characteristic time for a `sheet' of stabilizers to be evaluated. Obviously there is scope for far more sophisticated approaches which minimise waiting time, but this naive strategy suffices to give us a bound.
The $195$ steps required for $99\%$ completion are of course a mix of different operations: remote entanglement, local gates and measurement. From the physical timescales summarised in Supplementary Note 3 below, a picture emerges where $10\mu s$ may be a reasonable average time for an operation (assuming that there is progress on the crucial technical issues of photon loss, so that the more fundamental limits can be approached). We therefore estimate that $195$ steps will take approximately $2$ milliseconds. 

How severe will memory errors be in such a period? Experimentally reported memory qubit lifetimes have dramatically improved over the last couple of years, and one can hope that this may continue. As a note of caution we should however remember that our protocol will require many manipulations of other, nearby qubits while our memory qubits remain passive; memory performance in such a complicated environment has not been established. Maurer {\em et al} (Ref.~[23] of the main paper) reported that their best NV memory qubit had lifetimes of about 2 seconds (note the silicon impurity qubits studied by Steger {\em et al}  (Ref.~[24])  survived far longer, $\sim 3$ mins). Taking the shorter lifetime, if a proportion $1/e$ survive 2 seconds we infer an error rate of about $40\%$ per second, or a rate of  $0.1\%$ over our $2$ms protocol. This is nearly an order of magnitude below the error rates from the active processes, i.e. the gates and measurements, considered in the main paper. Obviously using the far longer lived silicon memory qubits would lead to an absolutely negligible error rate.

In comparison, the {\small STRINGENT} protocol takes about five times longer to achieve the same $99\%$ of complete stabilizers (see Supplementary Table S2). 
The minimum duration is 63 time steps,
which occurs with a probability 0.0422. Simulation of the distribution
found the mean duration of the stabilizer to be 278 time steps, while
50\% of operations were completed after 211 time steps, 95\% after
718 time steps, 99\% after 1067 time steps and 99.9\% after 1537 time
steps.
These higher costs in {\small STRINGENT} are the price paid for the increased error threshold as noted in the main paper. Whether {\small EXPEDIENT} or {\small STRINGENT} is the better protocol to adopt therefore depends on the relative severity of `active' errors associated with stabilizer measurement versus `passive' memory errors. }

\bigskip

{\color{black}
\noindent {\noindent {\bf \Large  Supplementary Note 3: Physical timescales \\}

Operations such as measurements, long range entanglement, and local qubit-qubit gates have timescales that vary considerably from one class of physical system to another. Moreover in some cases there is substantial potential for technological improvement, whereas other timescale are already close to fundamental limits. 

So called {\em single shot} optical measurement has now been accomplished in several classes of system (see e.g. PRL {\bf 100} 200502; {Science} {\bf 329} 599; {Nature} {\bf 467} 687). Typically there is a key period during which a driving laser gives rise to photon emission from the qubit system if, and only if, it is in a given state. Photon detector(s) monitor the qubit to observe such photos. The duration needs to be long enough that it is very likely at least one photon will be detected, if indeed the qubit is in the optically active state. Typical periods are of the order of $100\mu$s, although notably a $10\mu$s measurement achieving a fidelity beyond our requirements has been reported for trapped Ca ions (PRL {\bf 100} 200502). Also  a $5.5\mu s$ window has been employed in an NV system, albeit with limited fidelity ({Nature} {\bf 477} 574). In any case there is good scope to shorten the detection timescale in all systems suffering high rates of photon loss (a ubiquitous issue in current experiments) -- given more efficient detectors and less lossy optical interfaces, one may expect to see the measurement time fall below $10\mu$s.

The timescale for achieving remote entanglement can be very long in experimental demonstrations to date. This is true of both ion trap systems ({Nature} {\bf 449}, 68) and the very recent NV centre experiments (arXiv:1212.6136). The time requirements, which may extended to minutes, are again essentially due to the photon loss that occurs at various points in the system. Typically successful entanglement requires two specific photons to be detected, so that heralded failure occurs if {\em either} is lost. Therefore many such failures occur before eventual success.  Assuming that technical advances can largely remedy these losses, we can focus on a given instance of the underlying  entanglement protocol. The time for a single entanglement attempt can be dominated by the initialisation of the two qubits involved. In the work of Bernien {\em et al} (arXiv:1212.6136) this initialisation is composed of a pump phase followed by a measurement verification; the latter is affected by the photon loss issues described above, the former may be the more fundamental and was of $10\mu$s duration. In the atomic experiment of Moehring {\em et al} ({Nature} {\bf 449} 68) the time for a complete entanglement attempt was faster, around $2\mu$s.

Local conditional gates performed within a given structure typically rely on exploiting interactions between the two qubits involved, such as the hyperfine interaction between an electron and a nuclear spin. The magnitude of this interaction strength sets a limit on the speed with which a conditional evolution can take place. In practice other constraints such as the intensity of applied field may limit the speed further. Timescales reported in the experimental literature can vary from a few microseconds ({Science} {\bf 320}, 1326) to a few tens of microseconds ({Nature Physics} {\bf 208} 464).
}

\bigskip

\noindent {\noindent {\bf \Large   Supplementary Methods\\}

{\bf \noindent Twirling \\}
The protocols given in the main paper lead to slight irregularities between the weights associated with given errors occurring on different specific data qubits; so for example, in our superoperator the weight associated with $ZZ\Identity\Identity$ might be slightly higher or lower than the  weight associated with $\Identity ZZ\Identity$ . This is because the protocol performs $A-B$, $C-D$ pairings in Phase 1, and then $A-C$, $B-D$ pairing in Phase 2, thus the errors that survive this purification process will not be equally distributed under rotation of cell labels. While these irregularities have no significance for our threshold calculations, they do cause the weightings to have a spurious complexity. Therefore in our analysis we append onto the procotols of Fig.~2 a additional {\em twirling} operation which randomly applies swap operations between cells so as to `smooth' the weightings. Physically this is equivalent to the rather perverse  act of programming one's system with a range of possible protocols, identical except for permutation of the cell labels $A$ to $D$, and then applying one at random without retaining a record of the choice. But we emphasise that this is merely a theoretical convenience and there would be no need, or reason, to perform the process physically.

\bigskip

{\bf \noindent Example superoperators \\}
In the main paper we wrote the following to represent the overall effect of the stabilizer protocol on four data qubits initially in state $\rho$,
\[
P^M_{\small\rm real}(\rho)=\sum_e  a^M_eE_eP_{\small\rm ideal}^M(\rho)E_e^\dagger+b^{\bar M}_eE_eP_{\small\rm ideal}^{\bar M}(\rho) E_e^\dagger\ \ \ \tag{S1}
\label{superoperator}
\]
 \begin{equation}
 {\rm with}\ \ E_e=(ABCD)_e\ {\rm and}\ \{A,B,C,D\}\in\{\Identity,\sigma_x,\sigma_y,\sigma_z\}. \nonumber
 \end{equation}
Here $M$ stands for the reported outcome, ``odd'' or ``even'', and the $P_{\small\rm ideal}^M(\cdot)$ represents the perfect parity projector in either the $Z$ or $X$ basis depending on which class of stabilizer one is performing. The four operators making up $E$ are understood to act on data qubits 1 to 4 respectively, and index $e$ runs over all their combinations. The symbol ${\bar M}$ represents the compliment of $M$, i.e. ``odd'' for $M=$``even'' and vice versa. 

We noted that this real projector is made up of a mix of the `correct' and `incorrect' ideal projectors together with possible Pauli errors; the various weights $a$ and $b$ capture their relative significance. Here we give some examples of those weightings. The full list has 256 terms but they rapidly fall in magnitude so that the latter half are extremely small; here we list the terms corresponding to all single- and two-qubit errors. For the examples we will give, in fact the superscript $M$ can be omitted from the $a$ and $b$ weights, i.e. the same set of weights $a,b$ apply in $P_{\rm real}^{\rm even}$ as in $P_{\rm real}^{\rm odd}$. In other words the weight of the `correct' projector $P^M_{\rm real}$ is the same in each, as are the weights of all the various error combinations. 

In Supplementary Table~S3 the weights $A$ are associated with the `correct' projector $P^M_{\rm ideal}$ and the $B$ weights are associated with the `incorrect' projector $P^{\bar M}_{\rm ideal}$. The subscripts refer to the Pauli errors in the corresponding term, so that the subscript $\Identity$ indicates no errors, while subscript $X$ means ``a $\sigma_X$ on {\em one of} the four data qubits'', and $YZ$ means ``a $\sigma_Y$ on one of the qubits and a $\sigma_Z$ on another''. These capital letters $A$ and $B$ are therefore sum of one or more of the weights $a$, $b$ in Eqn.~\ref{superoperator}; for example $A_{X}=a_{\Identity\Identity\Identity X}+a_{\Identity\Identity X\Identity}+a_{\Identity X \Identity\Identity}+a_{X\Identity\Identity\Identity}$. As noted in a previous section, the twirling technique means that the individual $a$ or $b$ terms in such a sum in fact all have the same value.

Case 1 is the {\small EXPEDIENT} protocol operating with $p_n=0.1$ and $p_m=p_g=0.006$. Note that many of the corresponding $A$ and $B$ terms are identical due to symmetries in the algorithm (including the twirling). We see that there is a $91\%$ chance of a pure `correct' projection, a $6.2\%$ chance of a pure `wrong' projection, a $0.68\%$ chance of a correct projection followed by a single $\sigma_Z$ error, and so on.

Case 2 is {\small STRINGENT} operating with $p_n=0.1$ and $p_m=p_g=0.0075$. Finally for the comparison Case 3 is the {\em monolithic} circuit where there is no network noise to consider, and we set $p_m=p_g=0.09$.


\end{document}